\documentclass[a4paper]{jpconf}

\usepackage{amsmath}
\usepackage{amssymb}
\usepackage{graphicx}

\newcommand {\aplt} {\ {\raise-.5ex\hbox{$\buildrel<\over\sim$}}\ }

\begin{document}

\title{Research Update on Extreme-Mass-Ratio Inspirals}

\author{Pau Amaro-Seoane$^1$, Jonathan R. Gair$^2$, Adam Pound$^3$, Scott A. Hughes$^4$ and Carlos F. Sopuerta$^5$}

\address{$^1$ Max Planck Institut f\"ur Gravitationsphysik (Albert-Einstein-Institut), D-14476 Potsdam, Germany}
\address{$^2$ Institute of Astronomy, University of Cambridge, Madingley Road, Cambridge, CB3 0HA, UK}
\address{$^3$ Mathematical Sciences, University of Southampton, Southampton, SO17 1BJ, UK}
\address{$^4$ Department of Physics and MIT Kavli Institute, Massachusetts Institute of Technology, Cambridge, MA 02139, USA}
\address{$^5$ Institut de Ci\`encies de l'Espai (CSIC-IEEC), Campus UAB, Facultat de Ci\`encies, 08193 Bellaterra, Spain}

%\ead{Pau.Amaro-Seoane@aei.mpg.de,jgair@ast.cam.ac.uk,A.Pound@soton.ac.uk,sahughes@mit.edu,sopuerta@ieec.uab.es}
\ead{sopuerta@ieec.uab.es}

\begin{abstract}
The inspirals of stellar-mass mass compact objects into massive black holes in the centres of galaxies are one of the most important sources of gravitational radiation for space-based detectors like LISA or eLISA.  These extreme-mass-ratio inspirals (EMRIs) will enable an ambitious research program with implications for astrophysics, cosmology, and fundamental physics.
This article is a summary of the talks delivered at the plenary session on EMRIs at the 10th International LISA Symposium.  It contains research updates on the following topics: astrophysics of EMRIs; EMRI science potential; and EMRI modeling.
\end{abstract}

%%%%%%%%%%%%%%%%%%%%%%%%%%%%%%%%%%%%%%%%
% Introduction (Carlos F. Sopuerta)
%%%%%%%%%%%%%%%%%%%%%%%%%%%%%%%%%%%%%%%%
\section{Introduction}
One of the most important sources of gravitational waves (GWs) for a space-based detector like eLISA~\cite{eLISA:13,AmaroSeoane:2012je,eLISAGWN}
are extreme-mass-ratio binaries in the stage where the dynamics are driven by GW emission.  These systems, called EMRIs, are composed of a stellar-mass compact object (SCO) (a white dwarf, a neutron star, or a stellar-mass black hole (BH)) that inspirals into a massive
BH (MBH) located at a galactic centre.   The masses of interest for the SCO are in the range
$m = 1-10^2\; M^{}_{\odot}$, and for the MBH in the range $M = 10^5-10^7\; M_{\odot}$.
Then, the mass-ratio for these systems is in the interval $\mu=m/M \sim 10^{-7} -10^{-3}$.
The GW signals emitted by EMRIs are long lasting (months to years) and contain many GW cycles (of the order of $10^{5}$ during the last year before the SCO plunges into the MBH).  Many of these cycles are spent in the neighbourhood of the MBH horizon, meaning EMRI GWs encode a map of the strong-field region of the MBH.  These extraordinary features of EMRIs allow for a revolutionary research program, which could lead to understanding different aspects of stellar dynamics in galactic centres, tests of the geometry of BHs, and tests of General Relativity (GR) and alternative theories of gravity.  However, to implement such an ambitious program, several preliminary studies need to be done. In particular, we need to know how many events we expect to observe, which requires understanding which kind of astrophysical mechanisms can produce ERMI events.  Even more importantly, given the complexity of EMRI signals (which are essentially composed of a number of harmonics of three independent frequencies), we need a very precise theoretical
description of EMRI waveforms and also efficient data analysis search algorithms.  Finally, it is important to carry out theoretical studies in astrophysics, cosmology, and fundamental physics that will enable us to exploit all the scientific potential of EMRI GW observations.

This writeup offers a broad discussion of these themes, summarizing the three talks given in the plenary session on EMRIs~\cite{emrisession} at the 10th International LISA Symposium~\cite{lisax}, which was organized and chaired by Scott A. Hughes and Carlos F. Sopuerta.
The topics and organization of these talks are the following: the astrophysics of EMRIs (Pau Amaro-Seoane), in Sec.~\ref{emri-astrophysics}; an overview of EMRI science (Jonathan R. Gair), in Sec.~\ref{emri-scientificpotential}; and the status of EMRI modeling (Adam Pound), in Sec.~\ref{emri-modeling}.

%%%%%%%%%%%%%%%%%%%%%%%%%%%%%%%%%%%
% Pau Amaro-Seoane Contribution
%%%%%%%%%%%%%%%%%%%%%%%%%%%%%%%%%%%
\section{Astrophysics of EMRIs}\label{emri-astrophysics}

\subsection{Introduction}

One of the most exciting results of modern astronomy is the discovery, mostly
through high-resolution observations of the kinematics of stars and gas, that
most, if not all, nearby bright galaxies harbor a dark, massive, compact object
at their centers. The most spectacular case is our own
galaxy, the Milky Way.  By tracking and interpreting the stellar dynamics at
the centre of our galaxy, we have the most well-established evidence for the
existence of a MBH, with a mass of $\sim 4 \times 10^{6} \,
M_{\odot}$ \cite{EisenhauerEtAl05,GhezEtAl05,GhezEtAl08,GillessenEtAl09}. %\citep{EisenhauerEtAl05,GhezEtAl05,GhezEtAl08,GillessenEtAl09}.

These observations are difficult, especially for low-mass MBHs (ranging between
$10^5$ and $10^7\,M_{\odot}$) and even more so for the even less-massive
intermediate MBHs. Nowadays, using adaptive optics we could
optimistically hope to get a handful of measurements of stellar velocities of
such targets about $\sim 5$ kpc away in some ten years.  This is what makes an
experiment such as eLISA \cite{eLISA:13,AmaroSeoane:2012je,eLISAGWN} so %\citep{Amaro-SeoaneEtAl2013,Amaro-SeoaneEtAl2013b} so
appealing. Whilst main-sequence stars are tidally disrupted when approaching
the central MBH, SCOs slowly spiral into the MBH and are swallowed after some $\sim 10^{3-5}$
orbits in the eLISA band. At the closest approach to the MBH, the system emits
a burst of GWs which contains information about the spacetime geometry, the
masses of the system, and the spins of the MBH. We can regard each such burst
as a snapshot of the system.  By extracting the information encoded in the GWs
of this scenario, we can determine the redshifted mass of the central MBH with
an error of $\sim 1 - 0.1\,\%$ and for the spin of the MBH with $0.01\,\%$.

\subsection{The role of the MBH spin}

For a binary of an MBH and an SCO to be in the eLISA band, it has
to have a frequency of between roughly $10^{-4}$ and $1$ Hz. The emission of GWs
is more efficient as the SCO approaches its last stable orbit (LSO), so
that eLISA will detect the sources when they are close to the LSO line. The
total mass required to observe systems with frequencies between $10^{-4}$ and
$1$ Hz is $10^4 - 10^7\,M_{\odot}$. For masses larger than
$10^7\,M_{\odot}$, the frequencies close to the LSO will be too low, so
that their detection will be very difficult. On the other hand, for a total
mass of less than $10^3\,M_{\odot}$ we could in principal detect the EMRI at an
early stage, but then the amplitude of the GWs would be rather low.

Naively one could assume that the inspiral time is dominated by GW emission and
that if this is shorter than a Hubble time, the SCO will enter into an
EMRI with the MBH.  This is wrong, because one has to take into account the relaxation of
the stellar system.  Whilst it certainly can increase the eccentricity of the
SCO's orbit, it can also perturb the orbit and circularize it, so that the
inspiral time, $t^{}_{\rm GW}$, becomes larger than a Hubble time.
The condition for the SCO to be in an EMRI is that it is on an orbit
for which $t^{}_{\rm GW} \ll (1-e)\,t^{}_{\rm r}$ \cite{Amaro-SeoaneEtAl07}, with %\citep{Amaro-SeoaneEtAl07}, with
$t^{}_{\rm r}$ being the {\em local} relaxation time and $e$ the orbital eccentricity. When the orbit has a
semi-major axis for which the condition is not fulfilled, the SCO will have to be already on a so-called ``plunging orbit'', with $e\ge
e_{\rm plunge} \equiv 1-4\,R^{}_{\rm Schw}/a^{}_{\rm plunge}$, where $R^{}_{\rm Schw}$ is the
Schwarzschild radius of the MBH (i.e., $R^{}_{\rm Schw} =
2GM/c^{2}$, with $M$ being the MBH mass), and $a^{}_{\rm plunge}$ is the semi-major axis.

It has been
claimed a number of times by different authors that these plunging orbits would result in too
short a burst of gravitational radiation, which could only be detected if it
originated in our own Galactic Center \cite{HopmanFreitagLarson07} because one needs a %\citep{HopmanFreitagLarson07} because one needs a
coherent integration of some few thousand repeated passages through the
periapsis in the eLISA bandwidth. Therefore, such plunging objects would be lost from the GW signal,
since they would be plunging directly through the horizon of the MBH and
only a final burst of GWs would be emitted, and (i) such a burst would be very
difficult to recover, since the very short signal would be buried in a sea of
instrumental and confusion noise, and (ii) the information contained in the
signal would be practically nil.  There has been some work on the detectability
of such bursts~\cite{RubboEtAl2006,HopmanFreitagLarson07,YunesEtAl2008}, but %\citep{RubboEtAl2006,HopmanFreitagLarson07,YunesEtAl2008}, but
they would only be detectable in our galaxy or in the close neighborhood, and
the event rates are rather low, even in the most optimistic scenarios.

However, this picture is based on orbits around a non-spinning, Schwarzschild MBH, and it is significantly altered when the spin of the MBH is accounted for. Reference~\cite{Amaro-SeoaneSopuertaFreitag2013} estimated the number of cycles that
certain EMRI orbital configurations that were thought to be plunging orbits (or
orbits with insufficiently many cycles) in the case of non-spinning MBHs can spend in
a frequency regime of $f\in [10^{-4},1]$ Hz during their last year(s) of
inspiral before plunging into a spinning MBH.  It is important to assess how many of
these EMRIs will have sufficient Signal-to-Noise Ratio (SNR) to be detectable.
Reference~\cite{Amaro-SeoaneSopuertaFreitag2013} found that (prograde) EMRIs in a very radial orbit (that would lead to a plunge in the case of a Schwarzschild MBH) actually spend  a
significant number of cycles near the LSO in the Kerr case, more than sufficient to be detectable with good
SNR.  The number of cycles has previously been associated with $N^{}_{\varphi}$ (the
number of times that the azimuthal angle $\varphi$ advances $2\pi$), which is
usual for binary systems.  However, as we mentioned above, the structure of
the waveforms from EMRIs is quite rich since they contain harmonics of three
different frequencies: $f_{\varphi}$, $f_{r}$ (associated with the time to go from apoapsis to periapsis and back), and
$f^{}_{\theta}$ (associated with the time to complete an oscillation around the orbital plane in the case of inclined orbits).  Therefore the waveforms have cycles associated with these
three frequencies, $(f^{}_{r},f^{}_{\theta},f^{}_{\varphi})$, which makes them
quite complex, and in principle this is good for detectability (assuming we have
the correct waveform templates). Moreover, these cycles happen just before
plunge and take place in the strong field region very near the MBH horizon.
So these cycles should contribute more to the SNR than cycles taking place
farther away from the MBH horizon.

More interestingly, very highly eccentric EMRIs have an impact on event rates.
One can prove that

\begin{eqnarray}
\frac{{{a}_{\rm EMRI}^{\rm Kerr}}}{{{a}_{\rm EMRI}^{\rm Schw}}} & = & {\cal W}^{\frac{-5}{6-2\gamma}}(\iota,\,s)\,,\\
\frac{{\dot{N}_{\rm EMRI}^{\rm Kerr}}}{{\dot{N}_{\rm EMRI}^{\rm Schw}}} & = & {\cal W}^{\frac{20\gamma-45}{12-4\gamma}} (\iota,\,s) \,,
\label{eq.NAEMRIW}
\end{eqnarray}

\noindent
where $a_{\rm EMRI}$ is the maximum radius within which we estimate the event rate $\dot{N}_{\rm EMRI}$, and the ``Schw"/``Kerr" labels refer to results for a non-spinning (Schwarzschild)/spinning (Kerr) MBH. In these expressions, ${\cal W}$ is a function that depends on $\iota$, the
inclination of the EMRI orbit, and on $s$, the MBH dimensionless spin parameter\footnote{For the derivation and some
examples of values for ${\cal W}$, we refer the reader to the work of
\cite{Amaro-SeoaneSopuertaFreitag2013}.}.  We also have assumed that the stellar BHs
distribute around the central MBH following a power-law cusp of exponent
$\gamma$; i.e., that the density profile follows $\rho \propto r^{-\gamma}$
within the region where the gravity of the MBH dominates over the gravity of the
stars, with $\gamma$ ranging between 1.75 and 2 for the heavy stellar
components
\cite{Peebles72,BW76,BW77,ASEtAl04,PretoMerrittSpurzem04,AlexanderHopman09,PretoAmaroSeoane10,Amaro-SeoanePreto11}.
%\citep{Peebles72,BW76,BW77,ASEtAl04,PretoMerrittSpurzem04,AlexanderHopman09,PretoAmaroSeoane10,Amaro-SeoanePreto11}.
See~\cite{Gurevich64} for an interesting first appearance of this
concept\footnote{The authors obtained a similar solution for how electrons
distribute around a positively charged Coulomb centre.}.

For instance, for a spin of $s=0.999$ and an inclination of $\iota = 0.4\,$rad,
it is estimated that ${\cal W}\sim 0.26$ and, thus, $\dot{N}_{\rm EMRI}^{\rm Kerr}
\sim 30$. That is, {\em the event rates are boosted by a factor of 30} in comparison
to a non-rotating MBH.

Very eccentric EMRIs spend enough cycles
inside the band of eLISA to be detectable.  We note here that whilst it is true that EMRIs very near the
new separatrix shifted by the spin effect will probably not contribute enough
cycles to be detected, it is equally true for the old separatrix
(Schwarzschild, without spin).  In this sense, it is found that the spin generically increases
 the number of cycles inside the band for prograde EMRIs in such a
way that EMRIs very near to the non-spin separatrix, which contributed few
cycles, become detectable EMRIs.  In summary, spin increases the area, in
configuration space, of detectable EMRIs.  One can thus predict that EMRIs will be
highly dominated by prograde orbits.

Moreover, because spin allows for stable orbits very near the horizon in the
prograde case, the contribution of each cycle to the SNR is significantly
bigger than each cycle of an EMRI around a non-spinning MBH.

It has also been shown that vectorial resonant relaxation will not be efficient enough to
change prograde orbits into retrograde orbits once GW evolution dominates (which
would make the EMRIs plunge instantaneously, as they would be in a non-allowed
region of phase space).

More remarkably, the very eccentric EMRIs we have described here are solely produced by
two-body relaxation, a chaotic process, and as such, they are ignorant of the Schwarschild
barrier~\cite{Merritt:2011ve,Brem:2012et}, a phenomenon that drastically suppresses EMRI rates close to the MBH.  While low-eccentricity EMRIs run into the problem of having to find a
way to traverse this barrier, very eccentric EMRIs do not. One can therefore predict that EMRI
rates will be dominated by high-eccentricity binaries, with the proviso that
the central MBH is described by the Kerr solution of GR.

\subsection{Conclusions}

The event rate of very eccentric EMRIs is much larger than that of low-eccentricity EMRIs, as a number of
different studies by different authors using different methods have found (although in earlier works very eccentric EMRIs were enviasaged to be direct plunges
since they only considered Schwarzschild MBHs).  Up to now,
spin effects of the central MBH have been neglected. Hence, the question
arises, whether a plunge is really a plunge when the central MBH is spinning.
This consideration has previously been ignored.

To estimate EMRI event rates, one needs to know whether the orbital
configuration of the SCO is stable or not, because this is the
kernel of the difference between an EMRI and a plunge.
Reference~\cite{Amaro-SeoaneSopuertaFreitag2013} takes into account the fact that the spin
brings the LSO much closer to the horizon in the case of prograde orbits
but it pushes it away for retrograde orbits.  Since the modifications
introduced by the spin are not symmetric with respect to the non-spinning case,
and they are more dramatic for prograde orbits, one can prove that the inclusion of
spin increases the number of EMRI events by a significant factor. The exact
factor of this enhancement depends on the spin, but the effect is already quite
important for spins around $s \sim 0.7$.

Reference~\cite{Amaro-SeoaneSopuertaFreitag2013} proves that these very eccentric EMRIs do
spend enough cycles inside the band of eLISA to be detectable.  Although EMRIs very near
the new separatrix shifted by the spin effect will probably not contribute enough cycles to be detected, it is
equally true for the old separatrix, without spin.  The conclusion is that the presence of spin increases the area,
in configuration space, of detectable EMRIs.  As a consequence, EMRIs will be highly dominated by prograde orbits.

It has also been shown that these new kind of EMRIs originate in a region of
phase-space such that they will be ignorant of the Schwarschild barrier~\cite{Merritt:2011ve,Brem:2012et}, which drastically reduces the rates for low eccentricity EMRIs. The
reason for this is that they are driven by two-body relaxation and not resonant
relaxation.  While the boost in EMRI rates due to resonant relaxation is
affected by the Schwarzschild barrier, so that low-eccentricity EMRIs run into
the problem of having to find a way to cross it, high-eccentricity EMRIs are
already in the right place and led by two-body relaxation, a chaotic process ignorant of secular effects. The barrier affects
the production of EMRIs via torques, but not two-body relaxation, which is the
mechanism producing the high-eccentricity EMRIs.  Moreover, because the supression in
the rates is severe for those EMRIs with semi-major axes with values
approximately $a \gtrapprox 0.03$ pc, it has been predicted that the rates will be
dominated by the kind of EMRIs described in Ref.~\cite{Amaro-SeoaneSopuertaFreitag2013}.

In short, the existence of these highly eccentric EMRIs is good news: They sample regions closer to the MBHs' event horizons, offer richer signal sources, are not suppressed by secular effects (unlike lower-eccentricity EMRIs), enhance the event rates, and are louder and, hence, probe larger horizon distances.

%%%%%%%%%%%%%%%%%%%%%%%%%%%%
% Jon Gair Contribution
%%%%%%%%%%%%%%%%%%%%%%%%%%%%
\section{The scientific potential of EMRI observations} \label{emri-scientificpotential}
It is expected that a space-based gravitational wave detector similar to eLISA~\cite{eLISA:13} will observe a few tens of EMRI events per year~\cite{eLISAGWN}, although this number is subject to large astrophysical uncertainties. These events will be seen out to redshifts $z \sim 0.7$ and they will primarily be the inspirals of stellar BHs with mass $m \sim 10M_\odot$ into MBHs with mass $M\sim10^5$--$10^6M_\odot$~\cite{gair2009}. For every EMRI observed, eLISA will measure the system parameters to high precision, with typical fractional errors on masses of $\sim 10^{-4}$, typical errors on the spin of the central black hole of a few times $10^{-4}$ and typical precision on the sky location and distance of a source of a few square degrees and $\sim10\%$ respectively. These precise measurements come from accurately tracking the phase of the inspiral over the $\sim10^5$ waveform cycles observed for a typical system and are therefore not very dependent on the final design of the detector. The precise measurements for tens of systems that eLISA EMRIs will provide have tremendous potential to inform us about astrophysics, cosmology, and fundamental physics.

\subsection{Astrophysics}
Massive black holes in the low-mass range that will be observed by eLISA are not well constrained by electromagnetic observations, and there are therefore big uncertainties in the demographics of this population. EMRI observations will measure the mass and spin distributions of these MBHs in the local Universe for the first time. Using a simple power law model for the BH mass function, $dn/d\log M = AM^\alpha$, it was shown that with $N$ eLISA observations of EMRIs, constraints would be placed on the BH mass function at the level~\cite{gtv}
\begin{equation}
\Delta (\ln A) \approx 1.1 \sqrt{10/N}, \qquad \Delta(\alpha) \approx 0.35 \sqrt{10/N} .
\end{equation}
The current uncertainty in the slope of the mass function is $\sim \pm 0.3$, and so eLISA will improve on this with as few as $10$ EMRI observations. Predicted event rates are a factor of several larger than this. eLISA observations of EMRIs by themselves will not be able to probe evolution in the mass function with redshift~\cite{gtv}. The analysis in~\cite{gtv} assumed that the scaling of the astrophysical EMRI rate with the mass of the central MBH was known, so these constraints are really on the slope of the EMRI-rate-weighted mass function. While analytic and numerical calculations might improve our understanding of the rate scaling before eLISA is launched, the best chance to break this degeneracy is probably to use both eLISA observations of EMRIs and those of MBH mergers together to probe the mass function and its evolution.

EMRI observations have the potential to inform us about several other aspects of astrophysics, although these have not yet been investigated in detail. The properties of EMRI orbits measured by eLISA encode details of EMRI formation mechanisms. In the standard capture scenario, EMRIs would be expected to be on eccentric orbits inclined to the equatorial plane of the MBH, but other formation scenarios make different predictions: EMRI formation by the splitting of a binary would tend to create circular but inclined EMRIs; massive star formation in a disc would lead to circular and equatorial EMRIs; and formation of EMRIs via the tidal stripping of massive stars would lead to circular, inclined EMRIs involving low mass compact objects (white dwarfs)~\cite{Amaro-SeoaneEtAl07}. eLISA observations will indicate the relative importance of these different mechanisms. In addition, the rates, orbital properties and compact object masses of EMRI events probe the dynamical processes, such as relaxation and mass segregation, that drive them and the properties of stellar populations (e.g., mass functions) in galactic centres.

\subsection{Cosmology}
The amplitude of a gravitational wave source scales as $(1+z)M/D_L(z)$, where $M$ is the mass, $z$ the redshift and $D_L(z)$ the luminosity of the source. The redshifted mass, $(1+z)M$, can be determined very precisely from the observed phase evolution of the source, so provided that the mass-redshift degeneracy can be broken, gravitational wave sources can be used as ``standard sirens'' to measure $D_L(z)$ and probe the cosmological expansion history of the Universe~\cite{schutzH0}. Redshifts can be measured directly if an electromagnetic counterpart to a GW source is observed. For EMRIs, the only possible counterpart would be the tidal disruption of a white dwarf by a low-mass, highly-spinning BH~\cite{EMRIcounterpart}. However, it is extremely unlikely that any such EMRIs will be observed~\cite{gair2009}. Constraints can also be derived statistically using many EMRI events and this was examined in~\cite{McLeodHogan}. For each event, every observed galaxy consistent with the error bars from the GW measurement can be used to estimate the redshift of the source and hence a value for the luminosity distance or cosmological parameters. Averaging over multiple EMRI events provides final constraints on the cosmology. It was found that LISA observations of $\sim20$ events at redshift $z \aplt 0.5$ would be sufficient to measure the Hubble constant, $H_0$, to $\sim1\%$. eLISA will have slightly larger error bars for each EMRI, and therefore more events will be required to reach the same precision. However, we expect eLISA to observe several tens of EMRIs, all in this redshift range, and so we should be able to place constraints on $H_0$ at the level of $1-2\%$. Although electromagnetic constraints are likely to be more precise than this before eLISA is launched, the GW derived constraints will be completely independent, providing an important verification of existing results and a way to identify previously unknown systematics.

\subsection{Fundamental physics}
GW observations will probe sources in a regime of strong-field, non-linear and dynamical gravity that has never been explored, thus providing new tests of our understanding of the theory of gravity. All GW sources and detectors can be used for tests of fundamental physics, but EMRIs will provide particularly powerful probes as they will generate many tens of thousands of observable cycles in the strong-field regime, they are ``clean'' systems, as we expect most EMRIs to be BH binaries, and the orbital dynamics are rich---we expect most EMRIs to be both eccentric and inclined, so the emitted waveforms comprise a complex superposition of three fundamental frequencies---and the inspiralling object explores a large fraction of the space-time before it plunges. For a thorough review of the tests of fundamental physics that will be possible using EMRIs and other eLISA sources, we refer the interested reader to~\cite{TestGRLRR}, but we will summarise some of the main ideas here.

In general relativity, with some additional physicality assumptions, it is known that the Kerr BH is the unique end state of gravitational collapse~\cite{hawkingellis}. This is often referred to as the ``no-hair'' theorem, since the Kerr metric depends on only two parameters, mass $M$ and spin $a$, and all higher mass ($M_l$) and spin ($S_l$) multipole moments of the space-time are related to these as $M_l + iS_l = M (ia)^l$. The GWs emitted by an EMRI encode the orbital dynamics of the inspiralling object, which are in turn determined by the space-time structure. EMRI GWs thus encode a map of the space-time structure, which can be used to verify the no-hair property of the central black hole. This was first discussed in~\cite{ryan95}, in which a space-time with arbitrary multipole moments was used to show that the multipoles were separately encoded in gravitational wave observables, specifically the orbital precession rate and inspiral rate. The multipole expansion is not particularly practical, as an infinite number of multipoles are needed to represent Kerr. Recent work has instead focussed on ``bumpy'' black holes~\cite{CH2004}, in which the metric is Kerr plus a small perturbation. There have been many different studies of EMRI constraints on bumpy black holes~\cite{CH2004,GB2006,BCbumpy,GLM2008}, including BH solutions motivated by alternative theories of gravity~\cite{SY2009,CGS2012}. These studies conclude that EMRI observations will be able to simultaneously measure the MBH mass and spin to $\sim 0.01\%$ and a deviation in the quadrupole moment of the black hole of $\sim 0.1\%$. In addition, EMRI observations could place a bound on the deviation parameter characterising dynamical Chern Simons modified gravity (a parity-violating alternative to GR inspired by string theory) of $\xi^{\frac{1}{4}} < 10^4 {\rm km}$~\cite{CGS2012}.

GWs from EMRIs also encode qualitative information about the surface of the central object, whether or not the external space-time differs from Kerr. The presence/absence of a horizon would be indicated by a cut-off/continuation of GW emission after plunge (e.g., persistent emission for a massive boson star~\cite{KGK2005}); the properties of the surface of the central object can be inferred by comparing the observed inspiral rate and observed energy flux to deduce the energy lost to tidal interaction between the two objects~\cite{LiLove2008}, with $O(100\%)$ changes in the tidal energy loss being detectable with eLISA; and if the quasi-normal mode (QNM) spectrum of the central object is sufficiently different from that of Kerr, an EMRI could tidally excite QNMs as its orbital frequency passes through the QNM frequencies during an inspiral, leading to a resonant signature in the emitted GWs---e.g., for orbits around ``gravastars'', which comprise a thin shell of material with a de Sitter interior and Schwarzschild black hole exterior~\cite{pani2009}.

Measurable departures of an EMRI signal from the predictions of GR could also arise from astrophysical perturbations. The gravitational effect of a sufficiently dense and massive accretion torus could measurably perturb an EMRI~\cite{barausseA}, although this imprint can be mimicked by an error in the MBH mass and spin. In addition, hydrodynamic drag from material that the EMRI is moving through will qualitatively change the inspiral, by driving the orbit to become more prograde, while a GW driven inspiral should become more retrograde~\cite{barausseB}. However, in both cases the imprints will only be measurable if the amount of matter close to the central BH is unphysically high. If the inspiralling object in an EMRI is migrating through a massive disc a $\sim1$ radian dephasing could accumulate over a typical observation~\cite{Yunesdisc}. Finally, the presence of a second MBH within $\sim 0.1$pc would leave a detectable imprint~\cite{YMT2011}, while the presence of a second small object near the EMRI could lead to chaotic motion for $\sim 1\%$ of EMRIs~\cite{butterfly}. For each of these mechanisms, the number of systems in which a detectable signature could accumulate is small and, overall, we expect that these effects should leave a measurable imprint in only a small fraction ($\aplt 10\%$) of observed EMRIs. These imprints will be qualitatively different from fundamental physics effects and so will not be misinterpreted as the latter, but instead will provide constraints on the underlying astrophysical process.

EMRI observations can also be used in several other ways to test fundamental physics. Kerr orbits have a complete set of constants of motion, allowing the orbital motion to be separated, but this will not be true in generic space times, leading to qualitative differences in the dynamics that are potentially observable, such as the appearance of chaos late in the inspiral~\cite{GLM2008} or the appearance of persistent resonances during the inspiral~\cite{persistres}. EMRI observations can be used to test specific alternative theories of gravity (e.g., scalar-tensor gravity~\cite{scharre02,berti2005}, dynamical Chern Simons modified gravity~\cite{CGS2012} and scalar Gauss-Bonnet gravity~\cite{yagi2012}) or to place constraints on generic deviations from GR using phenomenological models~\cite{ppE,GY2011}. In most cases, the constraints possible with EMRIs are one or more orders of magnitude better than those that are available currently or likely to be obtained in the near future. EMRIs can be used to test that energy losses to gravitational waves are consistent with the quadrupole formula by comparing the inspiral rate to the quadrupole prediction. Indeed, the constraints on scalar-tensor gravity come primarily from detecting the excess energy lost to dipole radiation~\cite{scharre02}. Finally, EMRI observations can be used to test that GW polarisation is consistent with the two transverse-tensor modes expected in GR. Up to four alternative polarisation states exist in alternative theories of gravity, and eLISA's sensitivity to longitudinal modes is ten times greater than that to transverse modes at high frequencies~\cite{tinto2010}. Whether or not such modes are detectable will of course depend on the relative amplitude of excitement of the different modes in astrophysical sources. For a detailed discussion of all of these possibilities, see~\cite{TestGRLRR} and references therein.

%%%%%%%%%%%%%%%%%%%%%%%%%%%%%%
% Adam Pound Contribution
%%%%%%%%%%%%%%%%%%%%%%%%%%%%%%
\section{Status of EMRI modeling} \label{emri-modeling}

\subsection{Basic features and challenges}

Realizing the scientific potential of EMRI observations will require an accurate model of their waveforms. This poses many distinct challenges. The orbital motion is highly relativistic and the fields are strong, meaning post-Newtonian approximations are inapplicable. The system contains disparate lengthscales, the SCO's size $\sim m$ and the orbital scale $\sim M$, meaning full numerical relativity is unsuitable. Furthermore, the system contains disparate timescales: in addition to the orbital period $\sim M$, there is the inspiral time $t_{GW}\sim M/\mu$. For an astrophysically relevant mass ratio $\mu\sim 10^{-5}$, the inspiral time corresponds to an enormous $\sim 10^5$ wavecycles. Hence, to extract system parameters from an observed signal, a model must accurately match the true waveform's phase over this same, extraordinary number of orbits.

Despite these obstacles, substantial progress has been made in modeling EMRIs in recent years. Inspirals can now be simulated with an accuracy that may suffice for signal detection, if not for parameter estimation~\cite{Fujita-etal:09,Warburton-etal:12}. Numerous highly accurate calculations on the shorter, orbital timescale have been performed for generic orbits about a nonrotating BH~\cite{Barack-Sago:07,Detweiler:08,Barack-Sago:09,Barack-Sago:10,Barack-Sago:11}, as well as some for circular, equatorial orbits about a rotating one~\cite{Shah-etal:12,Dolan:13}. And the fundamental ingredients are now in place to simulate a complete inspiral accurately enough for parameter estimation, assuming the system is ``clean'' and described by classical GR~\cite{Hinderer-Flanagan:08, Pound:10a, Pound:12a, Gralla:12}. This progress has largely stemmed from advances in the self-force program~\cite{Barack:09,Poisson-Pound-Vega:11}, in which the SCO is treated as a source of perturbation of the background geometry of the MBH. Before detailing the various calculations that have been performed, we will briefly review the formalism.

%%%%%%%%%%%%%%%%%%%%%%%%%%%%%%%%%%%%%%%%
\subsection{The self-force formalism}
The gravitational self-force program, initiated nearly twenty years ago~\cite{Mino-Sasaki-Tanaka:97,Quinn-Wald:97}, is now mature and on firm mathematical footing~\cite{Gralla-Wald:08,Pound:10a,Harte:12}. Its essential idea is to write the full metric of the system, ${\sf g}_{\mu\nu}$, as an expansion
\begin{equation}\label{g_expansion}
{\sf g}_{\mu\nu} = g_{\mu\nu} + h^1_{\mu\nu} + h^2_{\mu\nu} + O(\mu^3),
\end{equation}
where $g_{\mu\nu}$ is the metric of the MBH, and $h^n_{\mu\nu}\sim \mu^n$ is the $n$th-order perturbation due to the SCO. The perturbations travel to infinity, where they can be observed as gravitational waves, but they also distort the geometry around the small object, influencing its motion; this effect on the motion is termed the self-force.

To cope with the challenges of EMRI modeling, this general picture must be tailored to accommodate both the multiple lengthscales and the multiple timescales in the problem. %The general strategy is to work outward from the small object, gaining knowledge of the object locally and then using that knowledge to formulate appropriate equations for the perturbations.

\subsubsection{Matched asymptotic expansions}\label{matching}

The multiple lengthscales are handled using the method of matched asymptotic expansions~\cite{Mino-Sasaki-Tanaka:97,Gralla-Wald:08,Pound:10a}. One first defines some appropriate worldline $z^\mu$ to represent the SCO's bulk motion in the background spacetime. Near the SCO, in a region of radius $r\sim m$ around $z^\mu$, the gravity of $m$ dominates over that of $M$, and the `outer expansion'~\eqref{g_expansion} fails to be accurate. In this region, a complementary `inner expansion' is assumed to exist, one that effectively zooms in on the SCO by letting $\mu\to 0$ while holding $r/m$ fixed. In this inner expansion, the background metric is that of the SCO itself. In a `buffer region', where $m\ll r\ll M$, both the inner and outer expansions are valid, and information about the SCO can be fed from the inner expansion to the outer. It turns out that in this procedure, only minimal information about the SCO is required---specifically, the SCO's multipole moments. An algorithm now exists for calculating the metric in the buffer region to arbitrary order in perturbation theory~\cite{Pound:12b}, and the end result is a local expression for each $h^n_{\mu\nu}$ in terms of the SCO's multipole moments.

\subsubsection{A generalized equivalence principle}
More precisely, the result of the local analysis in the buffer region is a split of $h^n_{\mu\nu}$ into two pieces:
\begin{equation}\label{split}
h^n_{\mu\nu}=h^{Sn}_{\mu\nu}+h^{Rn}_{\mu\nu}.
\end{equation}
The first piece, $h^{Sn}_{\mu\nu}$, loosely represents the SCO's bound field, being constructed locally from the SCO's multipole moments. The second piece, $h^{Rn}_{\mu\nu}$, propagates independently of the SCO and cannot be determined locally in the buffer region; it is determined only from the global boundary conditions imposed on the perturbations $h^n_{\mu\nu}$.

The fields $h^{Rn}_{\mu\nu}$ together form a smooth vacuum metric $\hat{\sf g}_{\mu\nu}=g_{\mu\nu}+\sum_n h^{Rn}_{\mu\nu}$, an effective metric in the neighbourhood of the SCO. This effective metric has proven to be extremely important: at least through second order in $\mu$, a sufficiently slowly spinning and sufficiently spherical\footnote{For a generic SCO, leading-order spin and quadrupole moment contributions to Eq.~\eqref{eq_mot} can be obtained straightforwardly~\cite{Gralla-Wald:08,Pound:10a,Pound:12b,Harte:12}, but order-$(m/M)^2$ subleading spin effects are unknown in the perturbative context~\cite{Pound:12b}.} SCO is known to move on a geodesic of $\hat{\sf g}_{\mu\nu}$~\cite{Pound:12a,Pound:14a}. In terms of the fields $h^{Rn}_{\mu\nu}$ on the background $g_{\mu\nu}$, the geodesic equation in $\hat{\sf g}_{\mu\nu}$ can be written as
\begin{align}\label{eq_mot}
\frac{D^2 z^\mu}{d\tau^2} &= \frac{1}{2}\left(g^{\mu\nu}+u^\mu u^\nu\right)\left(g_\nu{}^\rho-h^{R}_\nu{}^\rho\right)
								\left(h^{R}_{\sigma\lambda;\rho}-2h^{R}_{\rho\sigma;\lambda}\right)u^\sigma u^\lambda+O(\mu^3)\nonumber\\
								&\equiv F_1^\mu+F^\mu_2+O(\mu^3),
\end{align}
where $F_n^\mu$ is the $n$th-order self-force, $h^R_{\mu\nu}=h^{R1}_{\mu\nu}+h^{R2}_{\mu\nu}$, $\tau$ is proper time on $z^\mu$ as measured in $g_{\mu\nu}$, $u^\mu=\frac{dz^\mu}{d\tau}$, and a `;' denotes covariant differentiation compatible with $g_{\mu\nu}$. Equation~\eqref{eq_mot} is a statement of a \emph{generalized equivalence principle}: the SCO, regardless of its internal composition, falls freely in a certain  geometry $\hat{\sf g}_{\mu\nu}$, which in some sense it feels to be the ambient vacuum around it.

This generalized equivalence principle was first formulated at linear order in $\mu$ by Detweiler and Whiting \cite{Detweiler-Whiting:03}. A variant of it has been found to hold true even in the fully nonlinear, non-perturbative context of a material body with generic multipole structure~\cite{Harte:12}.

\subsubsection{Point particles and punctures}
To utilize the above results in a practical model, one must replace the inner expansion with something more useful in the region $r\sim m$. One does so by simply extending the result~\eqref{split} from the buffer region down to $r=0$. After this extension is performed, $h^{Sn}_{\mu\nu}$ gets called the singular field, because it diverges on $z^\mu$, while $h^{Rn}_{\mu\nu}$ gets called the regular field, because it is smooth at $z^\mu$. Crucially, this replacement of the physical metric in the SCO's interior with the extension from the buffer region does not alter the motion of the SCO or the metric outside the SCO.

At first order, the extended field $h^1_{\mu\nu}$ is identical to the perturbation produced by a point mass in linearized gravity~\cite{Gralla-Wald:08,Pound:10a}, meaning it can be calculated numerically from the linearized Einstein equation with a point-mass source. Beyond first order, the point particle picture fails, but one can still use this extension procedure to effectively replace the complicated physics inside the SCO with the simple field $h^{Sn}_{\mu\nu}$, which appears as a `puncture' (rather than a partcle) on the manifold. A globally well-defined field equation for the extended $h^n_{\mu\nu}$ is not known in this nonlinear case, but the field equations for $h^{Rn}_{\mu\nu}$ \footnote{More accurately, numerically practical field equations are formulated for `residual fields' $h^{\mathcal{R}n}_{\mu\nu}\approx h^{Rn}_{\mu\nu}$, which can be chosen such that their values on $z^\mu$, and the values of any finite number of their derivatives, agree with those of $h^{Rn}_{\mu\nu}$ and its derivatives on $z^\mu$.} in a region covering the SCO can easily be combined with equations for $h^n_{\mu\nu}$ outside the SCO to obtain a global solution~\cite{Detweiler:12,Pound:12a,Pound:12b,Gralla:12}.

\subsubsection{Representations of motion}\label{rep_motion}
In Sec.~\ref{matching}, we stated that ``One first defines some appropriate worldline $z^\mu$." That simple statement elides several complexities involved in the choice of worldline~\cite{Pound:10a}. In brief, on the orbital timescale $\sim M$, the orbit can be approximated as a geodesic of $g_{\mu\nu}$, and the perturbation can be calculated with the particle/puncture moving on that geodesic. Self-force effects then appear as order-$\mu$ corrections to the geodesic, and these corrections appear as terms in $h^2_{\mu\nu}$ rather than altering the position of the particle/puncture. But on the inspiral timescale $t_{GW}$, the orbit deviates by a large amount from any reference geodesic, and to maintain accuracy, the effect of the self-force must be incorporated into the motion of the particle/puncture.

There are two ways to incorporate this long-term backreaction of the perturbation on the worldline. First, one might solve the equation of motion~\eqref{eq_mot} coupled to the field equations for $h^n_{\mu\nu}$ and $h^{Rn}_{\mu\nu}$, with the particle/puncture moving on $z^\mu$; this is called the `self-consistent' approach~\cite{Gralla-Wald:08,Pound:10a}. Alternatively, on the right-hand side of Eq.~\eqref{eq_mot}, but not on the left, one might replace $z^\mu$ with an expansion $z^\mu=z_{0(\tau)}^\mu+z_{1(\tau)}^\mu+O(\mu^2)$, where $z^\mu_{0(\tau)}$ is the geodesic of $g_{\mu\nu}$ that is instantaneously tangential to $z^\mu$ at time $\tau$. At each instant $\tau$ on $z^\mu$, $F^\mu_1$ is then calculated from the solution to the linearized Einstein equation with the particle moving on $z^\mu_{0(\tau)}$, while $F^\mu_2$ picks up the effect of the correction $z^\mu_{1(\tau)}$. Call this the `geodesic source + osculation' approach, named after the fact that the source for $h^1_{\mu\nu}$ is at each instant on $z^\mu$ constructed from the osculating geodesic $z^\mu_{0(\tau)}$~\cite{Pound-Poisson:08,Warburton-etal:12}.

%%%%%%%%%%%%%%%%%%%%%%%%%%%%%%%%%%%%%%%%
\subsection{Self-forced motion in Kerr spacetime: A hierarchy of models}\label{models}
The self-force formalism can, in principle, be carried to any order in its perturbative expansion. But how high must we go to build useful EMRI models?

Geodesics in Kerr are characterized by their energy $E$, angular momentum $L_z$, and Carter constant $Q$, or equivalently, by their three orbital frequencies $f^{}_r$, $f^{}_\theta$, and $f^{}_\phi$. When the self-force is accounted for, its dissipative piece causes slow changes in these `constants' of motion. However, the conservative piece of the self-force also causes long-term changes that cannot be neglected~\cite{Pound-Poisson:08}. The relative importance of the various pieces of the self-force, through second order, was first systematically studied by Hinderer and Flanagan~\cite{Hinderer-Flanagan:08}. They used a two-timescale expansion of accelerated motion in Kerr to study generic behavior on both the orbital and inspiral timescales, leading to the following expansion of the orbital phase:
\begin{equation}
\phi(t,\mu) = \frac{1}{\mu}\left[\phi_0(\tilde t)+\mu\phi_1(\tilde t)+O(\mu^2)\right]
\end{equation}
where $\tilde t\equiv \mu t$ is $\sim M$ when $t\sim t_{GW}$. In this expansion, the leading term, $\phi_0$, called the adiabatic order, is determined by the averaged dissipative piece of $F_1^\mu$; at this order, only the leading averaged rates of change of $E$, $L_z$, and $Q$ are required. The subleading term, $\phi_1$, called the post-1 adiabatic order, is determined by the averaged dissipative piece of $F_2^\mu$, the conservative piece of $F_1^\mu$, and the oscillatory dissipative piece of $F_1^\mu$. A model that captures only $\phi_0$ is likely to suffice for signal detection but not for parameter estimation. A model that captures both $\phi_0$ and $\phi_1$ ensures all errors are small, $\sim \mu$, over an inspiral; therefore, it is expected to be sufficient for parameter estimation.

This description assumes the orbital frequencies $f^{}_r$ and $f^{}_\theta$ are incommensurate. When the frequencies are related by a rational ratio, orbital resonances occur and substantially complicate the picture. Because the orbital frequencies continually evolve, the resonance will last only a relatively short time $\sim M/\sqrt{\mu}\ll t_{GW}$.\footnote{Although it is possible for an orbit to become trapped near a resonance, such an event is extremely unlikely to be observed~\cite{vandeMeent:14}.} But during this time, the orbital elements change rapidly, leading to large, $\sim\sqrt{1/\mu}$ shifts in the waveform's phase~\cite{Flanagan-Hinderer:12}. These transient resonances occur generically in most inspirals~\cite{Ruangsri-Hughes:13}. They have interesting mathematical properties~\cite{Gair-etal:12} and observational relevance even outside of gravitational wave astronomy~\cite{Brink-etal:13}, but for the purpose of EMRI modeling, they are of great importance for one reason: even for signal detection, their effects might have to be included, but a leading-order adiabatic evolution does not capture those effects. Correctly modeling resonances is thought to have the same requirements as a post-1 adiabatic approximation~\cite{Flanagan-Hinderer:12}---the complete $F^\mu_1$ and the averaged dissipative effect of $F^\mu_2$. However, for detection, it might instead suffice to use the adiabatic approximation in the sections of waveform between resonances.

We now detail the calculations that have been performed, moving in order of increasing completeness, from adiabatic evolution to the full $F_1^\mu$ to $F^\mu_2$.

\subsubsection{Adiabatic evolution}

Building on seminal work by Mino~\cite{Mino:03}, several authors developed practical adiabatic approximation schemes that express the averaged rates of change $\langle\dot E\rangle$, $\langle\dot L_z\rangle$, and $\langle\dot Q\rangle$ in terms of the solution to the Teukolsky equation~\cite{Drasco-etal:05,Sago-etal:05,Sago-etal:06,Ganz-etal:07}. Compared to obtaining the complete $F^\mu_1$, these schemes are extremely simple, and they have been implemented for generic (inclined and eccentric) orbits in Kerr~\cite{Fujita-etal:09}. However, these calculations do not account for orbital resonances, possibly limiting their usefulness in signal detection. As a first step to overcoming this limitation, the effects of resonances on $\langle\dot E\rangle$, $\langle\dot{L}_z\rangle$, and $\langle\dot Q\rangle$ have been calculated~\cite{Flanagan-etal:12,Isoyama-etal:13}, and ``post-adiabatic'' methods for evolving through resonances are under development.

\subsubsection{First-order self-force}
There are now several active codes capable of calculating first-order effects in both the time domain \cite{Barack-Sago:07,Dolan-Barack:13} and the frequency domain \cite{Shah-etal:12,Akcay-Warburton-Barack:13,Osburn-etal:14}. One of two routes is typically used in these calculations. Either one calculates $h^1_{\mu\nu}$ using a point particle source and then subtracts the contribution of $h^{S1}_{\mu\nu}$ to find $h^{R1}_{\mu\nu}$ and $F_1^\mu$, as in mode-sum schemes~\cite{Barack-Ori:00,Barack-etal:02,Mino-Nakano-Sasaki:03,Barack-Ori:03b}; or one calculates the regular quantities $h^{R1}_{\mu\nu}$ and $F^\mu_1$ directly, as in puncture/effective-source schemes~\cite{Barack-Golbourn:07,Vega-Detweiler:07,Wardell-etal:11}.

Numerous calculations have been performed, and they are best divided into two categories: those for orbits about a Schwarzschild BH, and those for orbits about a Kerr BH. In all cases, calculations have used a geodesic source orbit, meaning they are accurate only on short timescales but can be incorporated into two-timescale or osculating-geodesic long-term evolution schemes.

For orbits in Schwarzschild, inspirals of moderate eccentricity have been simulated~\cite{Warburton-etal:12} using the `geodesic source + osculation' approach. Furthermore, many physical effects have been calculated from the conservative sector of the dynamics on the orbital timescale: orbital precession~\cite{Barack-Sago:11}, the shift in frequency of the innermost stable circular orbit (ISCO) in Schwarzschild~\cite{Barack-Sago:09,Favata:11,LeTiec-etal:12b}, Detweiler's redshift variable (the ratio of proper time measured along the geodesic in the regular metric to the time measured by an inertial observer at infinity)~\cite{Detweiler:08, Barack-Sago:11}, spin precession~\cite{Dolan-etal:13}, and tidal effects~\cite{Dolan-etal:14}.

For orbits in Kerr (the astrophysically relevant case), no inspirals have yet been simulated. However, calculations have been performed for circular orbits~\cite{Shah-etal:12,Dolan:13}, and from those results, Detweiler's redshift variable and the ISCO shift have been found~\cite{Isoyama-etal:14}. Also, recent work has established how to obtain self-force effects by reconstructing $h^1_{\mu\nu}$ from Teukolsky variables, rather than via the more technically demanding route of directly solving the nonseparable linearized Einstein equation~\cite{Shah-etal:12,Pound-Merlin-Barack:14}; this should spark a new round of calculations for more generic orbits.

%A first-order self-force approximation calculates the complete first-order metric perturbation $h^1_{\mu\nu}$, and from it calculates the complete first-order self-force. For the most part, these calculations have been performed using two methods:  and puncture/effective-source schemes . In mode-sum regularization, the tensor-harmonic modes of $h^1_{\mu\nu}$ are first calculated using a point-particle source, and then the harmonic modes of $h^{S1}_{\mu\nu}$ are substracted to obtain $h^{R1}_{\mu\nu}$ and the self-force. In a puncture scheme, the field equation is rewritten as an equation for $h^{R1}_{\mu\nu}$, which is then solved for directly; this is made possible by analytically approximating $h^{S1}_{\mu\nu}$ and moving it to the right-hand side of the field equation. Up to numerical errors, these two methods are exact and yield identical results.

Although short-term conservative effects in Schwarzschild and Kerr are not directly relevant to inspiral simulations, they have engendered a fruitful cross-cultural pollination between self-force models and other binary models~\cite{Detweiler:08,Blanchet-etal:10a,Blanchet-etal:10b,LeTiec-etal:11,Favata:11,LeTiec-etal:12b,LeTiec-etal:13,Shah-etal:13}, setting benchmarks for numerical relativity, fixing high-order PN parameters and calibrating effective-one-body theory (EOB)~\cite{Damour:09,Blanchet-etal:10b,Favata:11,Shah-etal:13, Barack-Damour-Sago:10,Barausse-etal:12,Akcay-etal:12,Bini-Damour:13}. More strikingly, the comparisons with numerical relativity have shown that self-force results are surprisingly accurate even in the limit of comparable masses~\cite{LeTiec-etal:11,LeTiec-etal:12b,LeTiec-etal:13}. See Ref.~\cite{LeTiec:14} for a recent review of this synergy between models.

All of the results mentioned in this section have been for the astrophysically relevant case of a gravitational perturbation. Significantly more progress has been made in the case of a scalar field, which has proven to be an important testbed for numerical methods in the self-force program~\cite{Detweiler-etal:03,Haas:07,Barack-Golbourn:07,Vega-Detweiler:07,Canizares-etal:10,Wardell-etal:14}. Fully self-consistent inspirals have been evolved in Schwarzschild in this case~\cite{Diener-etal:12}. And in the geodesic-source approximation, many calculations have been performed for orbits in Kerr~\cite{Warburton-Barack:10,Dolan-Barack-Wardell:11,Warburton-Barack:11}, leading to successful calculations of scalar self-force effects for both inclined~\cite{Warburton:14} and highly eccentric orbits in Kerr~\cite{Thornburg-Wardell:14}. Much of the numerical infrastructure developed for this work, and the physical insight gained from it, should carry over to the gravitational case.

\subsubsection{Second-order self-force}
After several preliminary studies of the second-order problem~\cite{Rosenthal:06a,Rosenthal:06b,Pound:10a,Detweiler:11}, in recent years complete second-order formalisms have been derived%, suitable for either Gralla-Wald or self-consistent representations of the motion
~\cite{Gralla:12,Pound:12a,Pound:12b,Pound:14a}. Because the point-particle description fails beyond linear order, mode-sum regularization can no longer be used. However, puncture schemes arise naturally. As of yet, no numerical implementations of these puncture schemes have been performed, but practical numerical ingredients and plans have been steadily developing~\cite{Pound-Miller:14,Pound:14c}.

\subsection{Summary of progress and current status}\label{status_summary}
The self-force formalism is now well developed and robust, extendable to any order in perturbation theory and to spinning objects of any multipolar structure.

To detect EMRI signals, the adiabatic approximation to the self-forced evolution is the most promising avenue. Excluding resonance effects, adiabatic evolutions have been simulated for generic orbits in Kerr~\cite{Fujita-etal:09}, and a method has been proposed to evolve through resonances without requiring a complete self-force calculation~\cite{Isoyama-etal:13}. However, while they are relatively cheap compared to a full self-force calculation, these adiabatic schemes may be too slow to be used directly to populate a template bank. For the purposes of testing data analysis methods and gleaning broad features of EMRIs, simple `kludge' models can be used instead~\cite{Barack-Cutler:04,Gair-Glampedakis:05,Babak-etal:07,Sopuerta-Yunes:11}. For building a template bank for real data analysis, one possible tool is the use of adiabatic approximations or full self-force data to calibrate the extreme-mass-ratio limit of EOB, which can then be used in EMRI source modeling~\cite{Damour-Nagar:07,Yunes-etal:09,Yunes-etal:11,Harms-etal:14}.

To accurately extract system parameters, we must improve upon these adiabatic models by including both the full first-order self-force and the time-averaged, dissipative effects of the second-order force. Progress toward this end has been steady, but is far from complete. Long-term evolutions incorporating the full first-order self-force have been performed for orbits of moderate eccentricity in a Schwarzschild background~\cite{Warburton-etal:12}. Short-term effects of the first-order self-force have been calculated for generic orbits in Schwarzschild~\cite{Barack-Sago:10} and for circular, equatorial orbits in Kerr~\cite{Shah-etal:12, Dolan:13}. And work has commenced to extend these results to generic orbits in Kerr and to incorporate them into orbital evolutions, either self-consistently or using a `geodesic source + osculation' approach.

The most daunting challenge now remaining is the incorporation of second-order self-force effects into inspiral models. This remains a distant prospect, and as of this writing, no numerical results have yet been obtained. However, the formalism is now at hand, and numerical calculations are underway~\cite{Pound:14c}. Furthermore, there is some evidence that for some types of orbits, particularly quasicircular ones, accurate models could be constructed using post-Newtonian approximations of second-order effects, potentially sidestepping the need for complete second-order calculations for those classes of orbits~\cite{Isoyama-etal:12}.

%%%%%%
\section{Conclusions}
We have summarized recent progress in three fundamental aspects of EMRI science: (i) the astrophysical mechanisms and event rates for EMRIs, (ii) the scientific potential of EMRI GW observations for astrophysics, cosmology, and fundamental physics, and (iii) the modeling of the dynamics of EMRIs and their GW signals.

Although there has been a lot of progress in these research topics, there is still a lot of work to be done in order to be ready for a fruitful exploitation of EMRI GW observations with a space-based detector like eLISA.  These three areas of research involve great theoretical challenges, which can be seen as part of the price to be paid for the extraordinary scientific results that can be obtained with EMRIs.  The case of EMRI modeling involves the solution of the general-relativistic two-body problem in the extreme-mass ratio regime, an interesting problem by itself.  As we have described, there is still much work to be done to solve the self-force problem for spinning MBHs, and even more for introducing second-order perturbative effects that are needed for an accurate description of EMRI gravitational waveforms.
In the case of the study of the astrophysical mechanisms and event rates for EMRIs, the astrophysical problem involves a huge range of scales that cannot be handled with direct $N$-body simulations, and clever approximations have to be made.  In the case of the scientific potential of EMRIs, there are a lot of possibilities, as we have described.  An example to illustrate the amount and complexity of the work that can be required to make progress in this direction is the question of testing alternative theories of gravity with EMRIs.  To that end, we would need a good description of EMRIs in those alternative theories of gravity.  Taking into account that the basic features of the GW emission mechanism and the properties of BHs may be quite different from those in GR, where EMRI dynamics is already quite challenging, making predictions for EMRI GW observations in alternative theories of gravity can be in general a very difficult problem.  However, by trying to solve this kind of problem we may learn new aspects of these alternative theories, especially in the strong-field gravity regime where not much is known.

In conclusion, the challenges originated by these questions about EMRIs and their intrinsic scientific interest require a great deal of human and computational resources.  Despite the fact that the L3 ESA mission devoted to the science of the Gravitational Universe white paper~\cite{eLISA:13} is expected to be launched around 2034, the problems we face are of sufficient magnitude to deserve increased attention in both the short and long term.

%%%%%%
\ack
PAS' work has been supported by the Transregio 7 ``Gravitational Wave Astronomy'' financed by the Deutsche Forschungsgemeinschaft DFG (German Research Foundation). AP acknowledges support from the European Research Council under the European Union's Seventh Framework Programme (FP7/2007-2013)/ERC grant agreement no. 304978. CFS acknowledges support from contracts 2009-SGR-935 (Catalan Agency for Research Funding, AGAUR) and AYA-2010-15709 (Spanish Ministry of Science and Innovation, MICINN).  SAH's work on EMRIs has most recently been supported in part by NSF grant PHY-1403261, by the John Simon Guggenheim Memorial Foundation, and with sabbatical support from the Canadian Institute for Theoretical Astrophysics and the Perimeter Institute for Theoretical Physics.

\section*{References}
\bibliographystyle{iopart-num}
%\bibliography{bibfile}

\providecommand{\newblock}{}

\end{document}